\documentclass[conference]{IEEEtran}
\usepackage[utf8]{inputenc}
\usepackage{CJKutf8}
\IEEEoverridecommandlockouts

\usepackage{cite}
\usepackage{amsmath,amssymb,amsfonts}
\usepackage{algorithmic}
\usepackage{graphicx}
\usepackage{textcomp}
\usepackage{xcolor}
\usepackage{multirow} 
\usepackage{booktabs} 
\usepackage{subfigure} 
\renewcommand{\arraystretch}{1} 

\def\BibTeX{{\rm B\kern-.05em{\sc i\kern-.025em b}\kern-.08em
    T\kern-.1667em\lower.7ex\hbox{E}\kern-.125emX}}
\begin{document}

\title{PGAD: Prototype-Guided Adaptive Distillation for Multi-Modal Learning in AD Diagnosis}

\author{
\IEEEauthorblockN{Yanfei Li\textsuperscript{1}, Teng Yin\textsuperscript{1}, Wenyi Shang\textsuperscript{1}, Jingyu Liu\textsuperscript{1}}
\IEEEauthorblockA{\textsuperscript{1}\textit{Machine Intelligence Lab, College of Computer Science and Technology} \\
\textit{Sichuan University}\\
Chengdu, China}
\and
\IEEEauthorblockN{Xi Wang\textsuperscript{2}}
\IEEEauthorblockA{\textsuperscript{2}\textit{Department of Computer Science and Engineering} \\
\textit{The Chinese University of Hong Kong}\\
Hong Kong, China}
\and
\IEEEauthorblockN{Kaiyang Zhao\textsuperscript{3}}
\IEEEauthorblockA{\textsuperscript{3}\textit{Department of Neurosurgery} \\
\textit{West China Hospital of Sichuan University}\\
Chengdu, China \\
zhanghaixian@scu.edu.cn}
}

\maketitle

\begin{abstract}
Missing modalities pose a major issue in Alzheimer’s Disease (AD) diagnosis, as many subjects lack full imaging data due to cost and clinical constraints. While multi-modal learning leverages complementary information, most existing methods train only on complete data, ignoring the large proportion of incomplete samples in real-world datasets like ADNI. This reduces the effective training set and limits the full use of valuable medical data. While some methods incorporate incomplete samples, they fail to effectively address inter-modal feature alignment and knowledge transfer challenges under high missing rates. To address this, we propose a Prototype-Guided Adaptive Distillation (PGAD) framework that directly incorporates incomplete multi-modal data into training. PGAD enhances missing modality representations through prototype matching and balances learning with a dynamic sampling strategy. We validate PGAD on the ADNI dataset with varying missing rates (20\%, 50\%, and 70\%) and demonstrate that it significantly outperforms state-of-the-art approaches. Ablation studies confirm the effectiveness of prototype matching and adaptive sampling, highlighting the potential of our framework for robust and scalable AD diagnosis in real-world clinical settings.
\end{abstract}


\section{Introduction}
Alzheimer’s Disease (AD) is a common neurodegenerative disorder that leads to irreversible cognitive decline, posing major challenges to global healthcare, especially as the population ages \cite{AlzReport2023,Weiner2015}. Early diagnosis, particularly at the Mild Cognitive Impairment (MCI) stage, is crucial for slowing disease progression \cite{Petersen2005,Jack2013,Gao1998}. Multi-modal neuroimaging, including Magnetic Resonance Imaging (MRI) and Positron Emission Tomography (PET), is widely used in AD research due to its ability to capture complementary structural and functional brain changes \cite{Zhang2023,Liu2022,Wang2024}. However, PET scans are expensive and not always accessible, leading to a significant portion of subjects in existing datasets lacking PET imaging. This reduces the availability of complete multi-modal data for AD studies \cite{Zuo2021}. As a result, many models cannot utilize data instances missing this modality, limiting their ability to fully exploit cross-modal information.

Most studies addressing this issue train on complete multi-modal data and later adapt to single-modality inference using techniques like knowledge distillation. This assumes full modality availability during training, making it less effective for real-world scenarios where missing data is common. Only a few works \cite{Kwak2024} have attempted to train directly on incomplete multi-modal data. However, without incorporating incomplete data into the teacher network’s training process, issues like unreliable inter-modal feature alignment \cite{Guan2022,Chen2023} and inefficient knowledge transfer \cite{Kwak2023,Wang2023} become more pronounced, reducing generalizability and diagnostic reliability.
To address this, we propose the PGAD framework, which integrates two key components: Prototype Consistency Matching (PCM) and Adaptive Multi-Modal Sampling (AMS). PCM reduces feature misalignment by learning class-level prototypes from complete multi-modal samples and enforcing consistency between missing modality features and their corresponding prototypes. This helps unpaired MRI samples retain essential modality-specific characteristics, minimizing feature discrepancies between missing and available modalities. Meanwhile, AMS improves the transfer of fused multi-modal features to MRI representations, ensuring that useful cross-modal information is effectively leveraged despite missing data.

In summary, our contributions are threefold: (1) We introduce PGAD, a novel knowledge distillation framework that directly incorporates incomplete multi-modal data into training, enhancing feature learning under missing conditions. (2) We propose PCM and AMS, which effectively mitigate modality gaps by aligning feature representations and dynamically balancing data utilization, improving model robustness. (3) We conduct extensive experiments on the ADNI dataset with varying missing rates (20\%, 50\%, and 70\%) and demonstrate that PGAD significantly outperforms existing approaches, achieving state-of-the-art performance in both AD classification and MCI conversion prediction.

\section{Method}
\begin{figure*}[t]
    \centering
    \includegraphics[width=\linewidth]{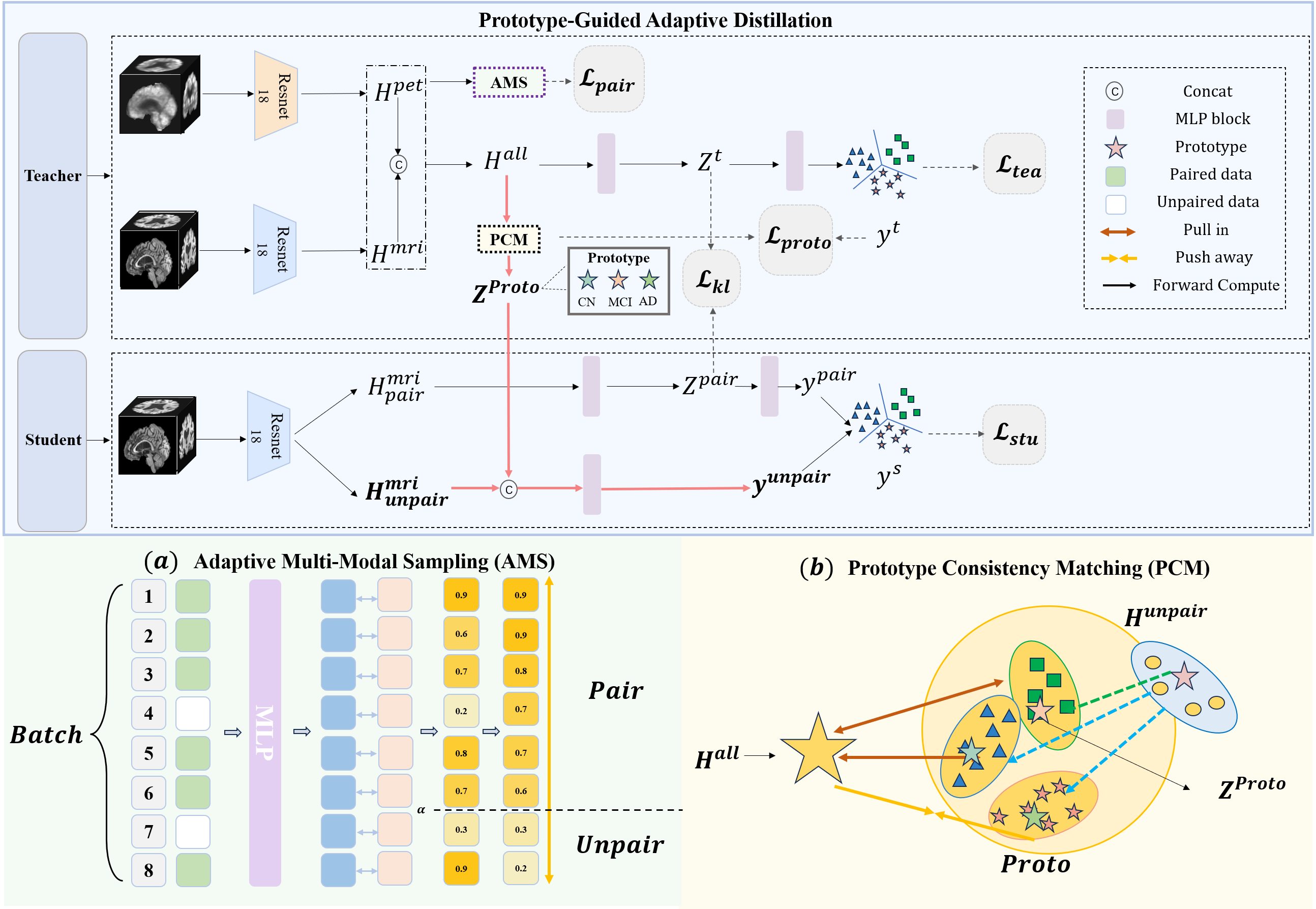}
    \caption{Overview of the PGAD framework. 
    The teacher network utilizes both MRI and PET data to extract joint feature representations and construct category prototypes. The student network, trained with only MRI data, operates under two conditions: (1) paired MRI samples ($H_{pair}^{mri}$) with corresponding PET data are constrained using distillation loss, and (2) unpaired MRI samples ($H_{unpair}^{mri}$) rely on category prototypes via the PCM module. The AMS module regulates the balance of paired and unpaired samples, enhancing training stability and cross-modal feature alignment.}
    \label{fig:PGAD_overview}
\end{figure*}

This study focuses on improving AD diagnosis under missing modality conditions by proposing the PGAD framework. PGAD is designed to enhance feature alignment, facilitate effective knowledge transfer, and stabilize training in the presence of incomplete multi-modal data. It achieves this by leveraging class prototypes to guide unpaired samples and employing a dynamic sampling strategy to regulate modality distributions. These mechanisms collectively improve diagnostic accuracy and model robustness. The overall framework is illustrated in Fig. \ref{fig:PGAD_overview}.
\subsection{PGAD Framework Overview}
The PGAD framework is designed to bridge the gap between complete and incomplete multi-modal learning by leveraging prototype-based feature alignment and adaptive sampling strategies. PGAD introduces two key components: PCM and AMS. PCM aligns unpaired MRI samples with category prototypes extracted from MRI-PET paired data, ensuring missing modality features retain meaningful representations. AMS dynamically adjusts the balance of paired and unpaired samples, mitigating data imbalance issues and improving cross-modal feature learning.

By combining prototype-guided knowledge transfer and structured sampling strategies, PGAD enables the student model to utilize available multi-modal information effectively while maintaining robustness when faced with missing modalities. This approach enhances feature consistency, stabilizes training, and improves generalization in real-world missing modality scenarios.

\subsection{Prototype Consistency Matching (PCM)}
In missing modality learning, complete data knowledge distillation relies on the teacher network's response to complete multi-modal inputs to guide the student network, while when the PET modality is missing, the teacher network cannot generate reliable and alignable outputs for the corresponding MRI sample, leading to the failure of the KL divergence loss function, and the student network struggles to learn cross-modal discriminative knowledge from the teacher network.

To address this critical issue, we propose the Prototype Consistency Matching (PCM) mechanism. Its core idea is to replace the instance-level teacher output with class-level prototype as the signal to guide the student network. Specifically, we leverage the fused features learned by the teacher network on complete multi-modal samples to construct a class prototype $z_c$ for each diagnostic category (e.g., AD, CN), which is the mean of the features of all complete samples of the same class:

\begin{equation}
    z_c = \frac{1}{N_c} \sum_{i \in C_c} h_i^{all},
\end{equation}

where $h_i^{all}$ is the joint feature extracted by the teacher network from a paired MRI-PET sample.

For an unpaired MRI sample $x_i^{\text{mri}}$, the student network extracts its feature representation $h_i^{\text{mri}}$. The PCM mechanism then computes the Euclidean distance between $h_i^{\text{mri}}$ and all class prototypes $z_c$, and assigns it to the nearest prototype $z_c^*$. The feature alignment is enforced by minimizing the squared Euclidean distance between $h_i^{\text{mri}}$ and $z_c^*$, which is formulated as the prototype consistency loss $L_{\text{proto}}$.

\begin{equation}
    L_{\text{proto}} = \frac{1}{|\mathcal{U}|} \sum_{i \in \mathcal{U}} \| h_i^{\text{mri}} - z_c^* \|^2,
\end{equation}

This design circumvents the limitations of KL divergence under missing modalities. The class prototype $z_c$, as a stable and consistent class center reference point, retains the discriminative information learned from complete multi-modal data. By pulling the features of unpaired samples towards these prototypes, PCM ensures that even in the absence of PET data, the MRI features learned by the student network remain semantically consistent with the complete multi-modal representation in the feature space, thereby enabling effective knowledge transfer for missing modalities.

\subsection{Adaptive Multi-Modal Sampling}
Conventional approaches typically confine teacher network training to complete multi-modal samples, leading to the underutilization of valuable incomplete data. This also presents a challenge when introducing our PCM module, which requires sufficiently representative class-level prototypes derived from fused multi-modal features.

To address this challenge, we propose a contrastive learning-based sampling mechanism that enables the teacher network to learn effectively from datasets containing both complete and incomplete samples. Our approach is grounded in a key insight: even when PET modality is missing in certain samples, the MRI data still contains valuable diagnostic information. The effective utilization of this information depends on the model's ability to understand the correspondence between MRI and PET modalities. To this end, we introduce a contrastive learning objective that enables the model to distinguish genuine MRI-PET paired samples from randomly combined unpaired samples.

\begin{equation}
    L_{\text{pair}} = -\frac{1}{|\mathcal{P}|} \sum_{(i,j) \in \mathcal{P}} \log \frac{\exp(\text{sim}(h_i^{\text{mri}}, h_j^{\text{pet}}))}{\sum_{k} \exp(\text{sim}(h_i^{\text{mri}}, h_k^{\text{pet}}))},
\end{equation}

In implementation, for each unpaired MRI sample $x_i^{\text{mri}}$ from class $c$, we construct a pseudo-paired sample by randomly selecting a PET scan $x_j^{\text{pet}}$ from the same class $c$ in the training set. This creates a mixed batch containing both genuine MRI-PET pairs and these synthetically generated pseudo-pairs. The contrastive learning objective then tasks the model with distinguishing between these two types of pairs.

In this framework, the positive pair $(i, j)$ denotes a genuine MRI-PET correspondence derived from a complete multi-modal input, while the negative samples $k$ comprise both randomly selected PET scans from other classes and those originating from pseudo-paired inputs. By encouraging the model to assign high similarity to true pairs and low similarity to pseudo or mismatched pairs, the teacher network is trained to capture authentic cross-modal associations. This discrimination ability is crucial for constructing a unified and semantically meaningful feature space, wherein the representations of unpaired MRI samples are effectively aligned with the multi-modal data manifold. Consequently, the class-level prototypes $z_c$ extracted from complete samples remain representative and informative, thereby enabling the PCM module to guide the learning of unpaired data more effectively.

To further refine this process, we introduce an adaptive mechanism for dynamically determining the optimal proportion of genuine paired samples to pseudo-paired samples within each training batch. Let $\theta$ denote a learnable scalar parameter. The actual sampling ratio $r$, which governs the fraction of genuine paired samples in the batch, is derived from $\theta$ via a sigmoid function, $r = \sigma(\theta)$, ensuring $r$ is bounded within the interval $[0, 1]$. This mechanism directly controls the composition of each training batch, with the remainder filled by pseudo-paired instances. Crucially, $\theta$ is not a fixed hyperparameter but a trainable parameter that is jointly optimized with the model's other parameters through standard backpropagation. The gradient signals derived from the contrastive loss $L_{\text{pair}}$ and the overall objective function guide the update of $\theta$, enabling the model to autonomously learn and adapt the sampling strategy to maximize feature alignment and learning efficiency throughout the training process.

This design not only circumvents the need for complex image generation processes inherent in traditional missing modality completion methods but also promotes stable feature learning and reliable diagnostic outcomes, thereby ensuring robust performance across diverse missing rate conditions. 

\subsection{Optimization Strategy}

The Prototype-Guided Adaptive Distillation (PGAD) framework is optimized through a combination of five loss terms that jointly enhance feature learning, knowledge transfer, and cross-modal alignment. The overall objective function is formulated as:
\begin{equation}
    L = \lambda_{\text{tea}} L_{\text{tea}} + \lambda_{\text{stu}} L_{\text{stu}} + \lambda_{\text{kl}} L_{\text{kl}} + \lambda_{\text{pair}} L_{\text{pair}} + \lambda_{\text{proto}} L_{\text{proto}}.
\end{equation}
The teacher classification loss $L_{\text{tea}}$ supervises the multi-modal teacher network to ensure effective feature extraction, while the student classification loss $L_{\text{stu}}$ guides the single-modality student model, enforcing consistency between paired and unpaired samples. The knowledge distillation loss $L_{\text{kl}}$ aligns the output distributions of the teacher and student models, enabling effective transfer of multi-modal knowledge. To enhance feature alignment, the paired sample consistency loss $L_{\text{pair}}$ constrains representations extracted from paired MRI and PET inputs, ensuring they retain modality-specific but complementary information. Finally, the prototype consistency loss $L_{\text{proto}}$ mitigates modality gaps by aligning unpaired MRI samples with category prototypes, preserving class-discriminative features and improving feature robustness. By integrating these five loss components, PGAD effectively balances classification, knowledge transfer, and feature consistency, ensuring stable learning under varying missing modality conditions and achieving superior performance in AD classification and MCI conversion prediction.

\section{Experiments} 
\subsection{Data Acquisition and Processing} 
\subsubsection{Dataset:}
This study utilized data from the Alzheimer’s Disease Neuroimaging Initiative (ADNI), including ADNI-1, ADNI-2, ADNI-GO, and ADNI-3. The selected samples all contain both T1-weighted MRI and FDG-PET modalities to ensure data completeness and consistency. The final dataset consists of 788 Alzheimer's Disease patients, 866 CN individuals, 370 pMCI patients, and 490 sMCI patients. The pMCI group includes individuals initially diagnosed with MCI who progressed to AD within 36 months, while the sMCI group comprises individuals who remained clinically stable as MCI for at least 36 months. Since specific age and gender information is not directly provided in the dataset, we estimated these attributes based on typical distributions in the ADNI database.

\setlength{\textfloatsep}{5pt} 
\setlength{\intextsep}{5pt} 
\setlength{\abovecaptionskip}{3pt} 
\setlength{\belowcaptionskip}{3pt} 

\begin{table}[ht]
    \centering
    \caption{Demographic information of the dataset}
    \label{tab:dataset}
    \begin{tabular}{lccc}
        \toprule
        \textbf{Class} & \textbf{Number} & \textbf{Gender (Male/Female)} & \textbf{Age (Mean ± SD)} \\
        \midrule
        AD   & 788 & 45\% / 55\% & 74 ± 8 \\
        CN   & 866 & 47\% / 53\% & 72 ± 7 \\
        pMCI & 370 & 46\% / 54\% & 73 ± 6 \\
        sMCI & 490 & 48\% / 52\% & 71 ± 7 \\
        \bottomrule
    \end{tabular}
\end{table}

\subsubsection{Preprocessing:}
 MRI preprocessing was performed using FSL, including reorientation, field of view estimation, skull stripping, and affine registration to the MNI152 template. PET preprocessing was conducted with SPM12, involving realignment, normalization, coregistration with MRI, image calculation, and smoothing. This ensures both modalities are spatially aligned for further analysis. 

\subsubsection{Experimental Settings:}
We evaluate the PGAD framework on Alzheimer’s Disease classification (AD vs. CN) and MCI conversion prediction (pMCI vs. sMCI). The dataset is split using stratified five-fold cross-validation, ensuring a balanced distribution of classes across folds. To simulate missing modalities, experiments are conducted under incomplete rates of 20\%, 50\%, and 70\%. During training, the teacher network processes both paired (MRI+PET) and unpaired (MRI with randomly assigned PET) data, while the student network is trained using only MRI. Final results are reported as the average performance across all five folds to ensure robustness and reliability.

The model is trained on four NVIDIA RTX 4090 GPUs with a batch size of 32 using the Adam optimizer (\(1 \times 10^{-4}\) initial learning rate, cosine annealing, weight decay \(5 \times 10^{-5}\)) for 100 epochs. The loss function consists of multiple terms with weights: \(\lambda_{\text{tea}} = 1.0\), \(\lambda_{\text{stu}} = 1.0\), \(\lambda_{\text{kl}} = 0.5\), \(\lambda_{\text{pair}} = 0.5\), and \(\lambda_{\text{proto}} = 0.5\).

\subsection{Comparison with Existing Approaches}
We compare PGAD with existing methods for Alzheimer's disease classification under AD vs. CN and pMCI vs. sMCI tasks. As shown in Table \ref{comparison_tab}, PGAD consistently achieves superior performance across all metrics, demonstrating the effectiveness of our prototype-guided adaptive distillation approach.

For fairness, PGAD is trained on complete multi-modal data and evaluated with MRI-only inputs, aligning with prior knowledge distillation-based methods. The performance of all compared methods was re-implemented and evaluated on our dataset under the same experimental protocol to ensure a consistent and fair comparison. All experiments are conducted using stratified five-fold cross-validation, and final results are reported as the mean performance with standard deviation across the five folds to quantify reliability.

In AD vs. CN classification, PGAD achieves the highest mean MCC (85.1±1.1) and AUC (96.3\%±0.4\%), demonstrating superior overall classification performance and consistency across folds. While DFTD shows competitive results with the second-best MCC (84.2±2.0) and AUC (96.2\%±0.5\%), PGAD maintains a more balanced performance profile with excellent mean sensitivity (93.4\%±1.2\%) that is critical for clinical diagnosis. 

For pMCI vs. sMCI prediction, PGAD achieves the highest mean MCC (71.3±2.0) and AUC (84.3\%±0.6\%), with particularly strong mean sensitivity (75.5\%±1.8\%) that is critical for early detection of progressive MCI cases. DFTD shows competitive performance with the second-best MCC (70.9±2.1) and AUC (83.9\%±0.7\%), but exhibits lower mean sensitivity (74.8\%±1.9\% vs. 75.5\%±1.8\%). The small standard deviations across folds for PGAD (MCC: ±1.1-2.0, AUC: ±0.4\%-0.6\%) indicate high reliability and robustness of our framework.

\setlength{\textfloatsep}{5pt} 
\setlength{\intextsep}{5pt} 
\setlength{\abovecaptionskip}{3pt} 
\setlength{\belowcaptionskip}{3pt} 
\begin{table}[htbp]
\centering
\caption{Comparisons of classification performance across different methods on AD vs. CN and pMCI vs. sMCI tasks. The best and second-best results in each metric are highlighted in bold and underlined. Values are reported as mean ± standard deviation over five-fold cross-validation.}
\label{comparison_tab}
\resizebox{\columnwidth}{!}{%
\begin{tabular}{l|l|cccc}
\toprule
\multirow{2}{*}{Task} & \multirow{2}{*}{Method} & \multicolumn{4}{c}{Metrics} \\
\cmidrule(lr){3-6}
 &  & AUC (\%) & MCC & SEN (\%) & SPE (\%) \\
\midrule
\multirow{7}{*}{AD vs. CN}  
 & 3D-Mixer \cite{zhang2023improving} & 94.7±0.8 & 76.1±3.2 & 90.5±1.6 & 91.3±1.4 \\
 & DA-MIDL \cite{zhu2021dual} & 95.4±0.7 & 80.2±2.1 & 91.7±1.5 & 92.2±1.3 \\
 & GF-NET \cite{Zhang2022} & 95.8±0.6 & 82.3±2.8 & 92.3±1.4 & 92.8±1.2 \\
 & DFTD \cite{Chen2023} & \underline{96.2±0.5} & \underline{84.2±2.0} & \underline{92.9±1.4} & \textbf{93.4±1.1} \\
 & Wang et al. \cite{Wang2024} & 95.2±0.7 & 79.1±3.1 & 91.3±1.6 & 91.8±1.5 \\
 & MDT-Student et al. \cite{Kwak2024} & 95.1±0.8 & 78.3±3.2 & 91.0±1.7 & 91.6±1.6 \\
 & \textbf{PGAD (Ours)} & \textbf{96.3±0.4} & \textbf{85.1±1.1} & \textbf{93.4±1.2} & \underline{93.2±1.3} \\
\midrule
\multirow{4}{*}{pMCI vs. sMCI}  
 & 3D-Mixer \cite{zhang2023improving} & 82.9±0.9 & 66.2±3.6 & 73.7±1.9 & 77.7±1.7 \\
 & DA-MIDL \cite{zhu2021dual} & 83.4±0.8 & 68.3±3.4 & 74.3±1.8 & 78.3±1.6 \\
 & DFTD \cite{Chen2023} & \underline{83.9±0.7} & \underline{70.9±2.1} & \underline{74.8±1.9} & \textbf{79.2±1.4} \\
 & \textbf{PGAD (Ours)} & \textbf{84.3±0.6} & \textbf{71.3±2.0} & \textbf{75.5±1.8} & \underline{79.0±1.5} \\
\bottomrule
\end{tabular}%
}
\end{table}

To further validate the statistical significance of PGAD's performance advantage, we conduct a comprehensive paired t-test analysis comparing PGAD against all competing methods on the AD vs. CN task. The p-values are calculated based on the performance scores across the five cross-validation folds, providing a rigorous assessment of whether the observed improvements are statistically meaningful. To enhance the rigor of the statistical testing and control the overall false positive rate, we apply the Bonferroni correction. The original significance level ($\alpha=0.05$) is divided by the total number of comparisons performed (6 methods $\times$ 4 metrics = 24), yielding a new, more stringent significance threshold of $\alpha'=0.00208$.

Table~\ref{tab:pvalue_analysis} presents the p-values for the paired t-test comparing PGAD against each competing method across all four evaluation metrics. The results show that PGAD's improvements over 3D-Mixer, DA-MIDL, GF-NET, and MDT-Student et al. are statistically significant in all metrics, with p-values well below the corrected threshold of 0.00208. Compared to Wang et al., PGAD achieves significant improvements in MCC (p=0.0018) and AUC (p=0.0022). Most notably, PGAD demonstrates a statistically significant superiority over the second-best method, DFTD, in sensitivity (p=0.0019), which is a critical metric for clinical diagnosis. While PGAD also outperforms DFTD in MCC (p=0.0021) and AUC (p=0.0024), the difference in specificity is not significant (p=0.072), indicating that DFTD's slight edge in specificity is likely within the expected variation. The low standard deviations for DFTD (e.g., MCC: ±2.0) in Table~\ref{comparison_tab} confirm its high stability, making the significant p-values against it a strong validation of PGAD's effectiveness.

\begin{table}[htbp]
\centering
\caption{P-value analysis for paired t-test comparing PGAD against existing methods on the AD vs. CN task. The significance level is set to $\alpha'=0.00208$ after Bonferroni correction.}
\label{tab:pvalue_analysis}
\begin{tabular}{lcccc}
\toprule
\multirow{2}{*}{Method} & \multicolumn{4}{c}{p-value} \\
\cmidrule(lr){2-5}
 & MCC & AUC (\%) & SEN (\%) & SPE (\%) \\
\midrule
3D-Mixer \cite{zhang2023improving} & \textless{}0.001 & \textless{}0.001 & \textless{}0.001 & \textless{}0.001 \\
DA-MIDL \cite{zhu2021dual} & \textless{}0.001 & \textless{}0.001 & \textless{}0.001 & \textless{}0.001 \\
GF-NET \cite{Zhang2022} & 0.0015 & 0.0017 & 0.0012 & 0.0019 \\
DFTD \cite{Chen2023} & 0.0021 & 0.0024 & 0.0019 & 0.072 \\
Wang et al. \cite{Wang2024} & 0.0018 & 0.0022 & 0.0025 & 0.065 \\
MDT-Student et al. \cite{Kwak2024} & \textless{}0.001 & \textless{}0.001 & \textless{}0.001 & \textless{}0.001 \\
\bottomrule
\end{tabular}
\end{table}

\subsection{Ablation Study}
We conduct a series of ablation studies to systematically evaluate the contributions of key components in our PGAD framework. These experiments are designed to validate the effectiveness of PCM and AMS mechanisms, determine optimal hyperparameters, and assess prototype construction strategies under varying missing modality conditions.

\subsubsection{Component Analysis: PCM and AMS}
To quantify the impact of our core components, we evaluate the framework's performance with and without PCM and AMS under different missing rates (20\%, 50\%, and 70\%). All experiments are conducted using stratified five-fold cross-validation, and results are reported as mean ± standard deviation to assess reliability. As shown in Table~\ref{ablation_study}, the absence of PET data significantly degrades performance compared to the full modality setting (MCC 85.1±1.1 vs. 80.2±2.1 at 20\% missing rate), highlighting the importance of multi-modal information.

The incorporation of PCM alone consistently improves MCC across all missing rates, with the most significant gains observed at 50\% missing rate (81.3 vs. 79.2, +2.1 MCC). This demonstrates PCM's effectiveness in aligning unpaired MRI features with class prototypes, preserving discriminative information despite missing modalities. When both PCM and AMS are integrated, the framework achieves optimal performance at lower missing rates (83.2 at 20\% missing rate), while maintaining robustness at higher missing rates (76.4 at 70\% missing rate). Notably, at 50\% missing rate, the full model achieves the highest sensitivity (94.0\%±1.3\%), indicating its superior ability to identify positive cases under challenging conditions. The consistently low standard deviations across all configurations (typically $<3.0$ for MCC and $<1.5\%$ for other metrics) confirm the stability and reliability of our framework.

\begin{table}[htbp]
\centering
\caption{Ablation study on the effect of PCM and AMS under different missing modality ratios. Values are reported as mean ± standard deviation over five-fold cross-validation.}
\label{ablation_study}
\setlength{\tabcolsep}{4pt}
\renewcommand{\arraystretch}{1.1}
\resizebox{\columnwidth}{!}{%
\begin{tabular}{c|cc|cccc}
\toprule
Missing Ratio & PCM & AMS & MCC & SEN (\%) & SPE (\%) & AUC (\%) \\
\midrule
Full PET & \checkmark & \checkmark & 85.1±1.1 & 93.4±1.2 & 93.2±1.3 & 96.3±0.4 \\
\midrule
\multirow{3}{*}{0.2} 
  & $\times$ & $\times$ & 80.2±2.1 & 93.8±1.5 & 91.5±1.4 & 95.0±0.7 \\
  & \checkmark & $\times$ & 82.3±2.8 & 94.5±1.4 & 92.2±1.3 & 95.3±0.6 \\
  & \checkmark & \checkmark & 83.2±2.2 & 95.8±1.1 & 92.8±1.3 & 95.8±0.5 \\
\midrule
\multirow{3}{*}{0.5} 
  & $\times$ & $\times$ & 79.2±2.7 & 94.8±1.3 & 91.2±1.5 & 94.2±0.8 \\
  & \checkmark & $\times$ & 81.3±2.1 & 93.5±1.5 & 93.8±1.2 & 95.0±0.7 \\
  & \checkmark & \checkmark & 82.4±1.3 & 94.0±1.3 & 92.4±1.4 & 94.6±0.6 \\
\midrule
\multirow{3}{*}{0.7} 
  & $\times$ & $\times$ & 73.1±2.8 & 92.5±1.7 & 87.0±1.8 & 90.5±0.9 \\
  & \checkmark & $\times$ & 74.2±2.6 & 92.8±1.6 & 87.8±1.7 & 91.0±0.8 \\
  & \checkmark & \checkmark & 76.4±1.1 & 91.5±1.5 & 90.5±1.6 & 92.3±0.7 \\
\bottomrule
\end{tabular}%
}
\end{table}

\subsubsection{Loss Function Hyperparameter Analysis}
To determine the optimal balance among loss components, we conducted a systematic study of hyperparameters $\lambda_{tea}$, $\lambda_{stu}$, $\lambda_{kl}$, $\lambda_{pair}$, and $\lambda_{proto}$ under 50\% missing modality rate. Starting with equal weights (all 1.0), we sequentially reduced individual weights while maintaining others constant. All experiments were performed using stratified five-fold cross-validation, and results are reported as mean ± standard deviation to assess reliability.

As presented in Table~\ref{tab:ablation_hyperparams}, the baseline configuration (all weights=1.0) achieves MCC 82.3±2.8 and AUC 95.5\%±0.8\%. Reducing $\lambda_{kl}$ to 0.5 improves AUC to 95.8\%±0.7\%, suggesting that excessive knowledge distillation pressure may hinder the student network's ability to learn modality-specific features. Similarly, reducing $\lambda_{pair}$ and $\lambda_{proto}$ to 0.5 individually yields improvements in MCC (83.5±2.1 and 83.2±2.0 respectively). The optimal configuration combines all three reductions ($\lambda_{kl}=\lambda_{pair}=\lambda_{proto}=0.5$), achieving MCC 85.1±1.1 and AUC 96.3\%±0.4\%. This indicates that while classification losses ($\lambda_{tea}$ and $\lambda_{stu}$) should maintain full weight to ensure task performance, the auxiliary losses benefit from slightly reduced influence to prevent over-constraining the optimization landscape. The consistently low standard deviations across all configurations confirm the stability of the hyperparameter selection process.

\begin{table}[t]
\centering
\caption{Ablation study on loss function hyperparameters for AD vs. CN classification under 50\% missing modality. Values are reported as mean ± standard deviation over five-fold cross-validation.}
\label{tab:ablation_hyperparams}
\setlength{\tabcolsep}{4pt}
\renewcommand{\arraystretch}{1.1}
\resizebox{\columnwidth}{!}{%
\begin{tabular}{c c c c c c c}
\toprule
$\lambda_{tea}$ & $\lambda_{stu}$ & $\lambda_{kl}$ & $\lambda_{pair}$ & $\lambda_{proto}$ & MCC & AUC(\%) \\
\midrule
1.0 & 1.0 & 1.0 & 1.0 & 1.0 & 82.3±2.8 & 95.5±0.8 \\
1.0 & 1.0 & 0.5 & 1.0 & 1.0 & 83.2±2.1 & 95.8±0.7 \\
1.0 & 1.0 & 1.0 & 0.5 & 1.0 & 83.5±2.1 & 95.9±0.6 \\
1.0 & 1.0 & 1.0 & 1.0 & 0.5 & 83.2±2.0 & 95.7±0.7 \\
\textbf{1.0} & \textbf{1.0} & \textbf{0.5} & \textbf{0.5} & \textbf{0.5} & \textbf{85.1±1.1} & \textbf{96.3±0.4} \\
\bottomrule
\end{tabular}%
}
\end{table}

\subsubsection{Prototype Construction Strategy Analysis}
To validate the effectiveness of our dynamic prototype construction approach, we compare three strategies for guiding unpaired MRI samples under 50\% missing modality rate: (1) No Prototype: Disabling PCM entirely, relying only on classification and distillation losses; (2)All Prototypes: Using global prototypes constructed from all available paired samples in the training set; (3)Paired Prototypes: Our proposed approach using batch-specific prototypes from currently available paired samples. All experiments are conducted using stratified five-fold cross-validation, and results are reported as mean ± standard deviation.

As shown in Table~\ref{tab:ablation_prototype_source}, the No Prototype baseline achieves MCC 78.2±2.7, confirming that explicit guidance for unpaired samples is essential for maintaining feature consistency. Using All Prototypes improves performance to MCC 82.1±2.0, demonstrating the value of prototype-based alignment in preserving class-discriminative information. However, our Paired Prototypes approach achieves the highest performance (MCC 85.1±1.1), indicating that contextually relevant, dynamically constructed prototypes provide superior guidance compared to static global prototypes. This is particularly evident in sensitivity (93.4\%±1.2\% vs. 91.9\%±1.4\%), suggesting that batch-specific prototypes better capture the nuanced feature relationships within each training iteration, leading to more robust classification of positive cases. The low standard deviation of our method (±1.1 for MCC) further confirms its reliability and stability.

\begin{table}[t]
\centering
\caption{Ablation study on prototype guidance strategies for AD vs. CN classification under 50\% missing modality. Values are reported as mean ± standard deviation over five-fold cross-validation.}
\label{tab:ablation_prototype_source}
\setlength{\tabcolsep}{5pt}
\renewcommand{\arraystretch}{1.1}
\resizebox{0.9\columnwidth}{!}{%
\begin{tabular}{l c c c c}
\toprule
Prototype Strategy & MCC & AUC(\%) & SEN(\%) & SPE(\%) \\
\midrule
No Prototype & 78.2±2.7 & 94.5±0.8 & 90.8±1.6 & 91.5±1.5 \\
All Prototypes & 82.1±2.0 & 95.2±0.7 & 91.9±1.4 & 92.4±1.3 \\
Paired Prototypes (Ours) & \textbf{85.1±1.1} & \textbf{96.3±0.4} & \textbf{93.4±1.2} & \textbf{93.2±1.3} \\
\bottomrule
\end{tabular}%
}
\end{table}

\subsubsection{Ablation Study on AMS Mechanism}
To evaluate the effectiveness of the dynamic sampling strategy in AMS, we compare three variants under 50\% missing rate: (1) No AMS: no sampling strategy applied; (2)Fixed-ratio AMS: a fixed 50\% ratio for paired/unpaired samples; (3)Dynamic-ratio AMS (Ours): our proposed adaptive mechanism.

As shown in Table~\ref{tab:ablation_ams}, Fixed-ratio AMS outperforms No AMS, demonstrating that balanced sampling is beneficial. Our Dynamic-ratio AMS achieves the best performance, validating its effectiveness. The small performance gap between Fixed-ratio and Dynamic-ratio AMS indicates that the dynamic mechanism provides a subtle yet consistent refinement over a well-chosen fixed ratio.

\begin{table}[htbp]
\centering
\caption{Ablation study on the AMS mechanism under 50\% missing rate. Results are reported as mean ± standard deviation over five-fold cross-validation.}
\label{tab:ablation_ams}
\begin{tabular}{lccc}
\toprule
Sampling Strategy & MCC & AUC (\%) & SEN (\%) \\
\midrule
No AMS & 80.2±2.1 & 94.5±0.7 & 92.8±1.5 \\
Fixed-ratio AMS & 83.2±2.2 & 95.2±0.6 & 94.3±1.2 \\
Dynamic-ratio AMS (Ours) & \textbf{85.1±1.1} & \textbf{96.3±0.4} & \textbf{93.4±1.2} \\
\bottomrule
\end{tabular}
\end{table}

\subsection{Visualization Analysis of Experimental Results}
This section presents a comprehensive analysis of the model's learned representations and attention mechanisms, providing insights into its robustness and clinical interpretability under missing modality conditions. We employ t-SNE for global feature space analysis and Grad-CAM for local biomarker identification.

\subsubsection{Feature Representation Analysis via t-SNE}
\begin{figure}[t]
\centering
\resizebox{\columnwidth}{!}{%
\begin{tabular}{cccc}
\subfigure[No PCM, No AMS]{
\includegraphics[width=0.29\linewidth]{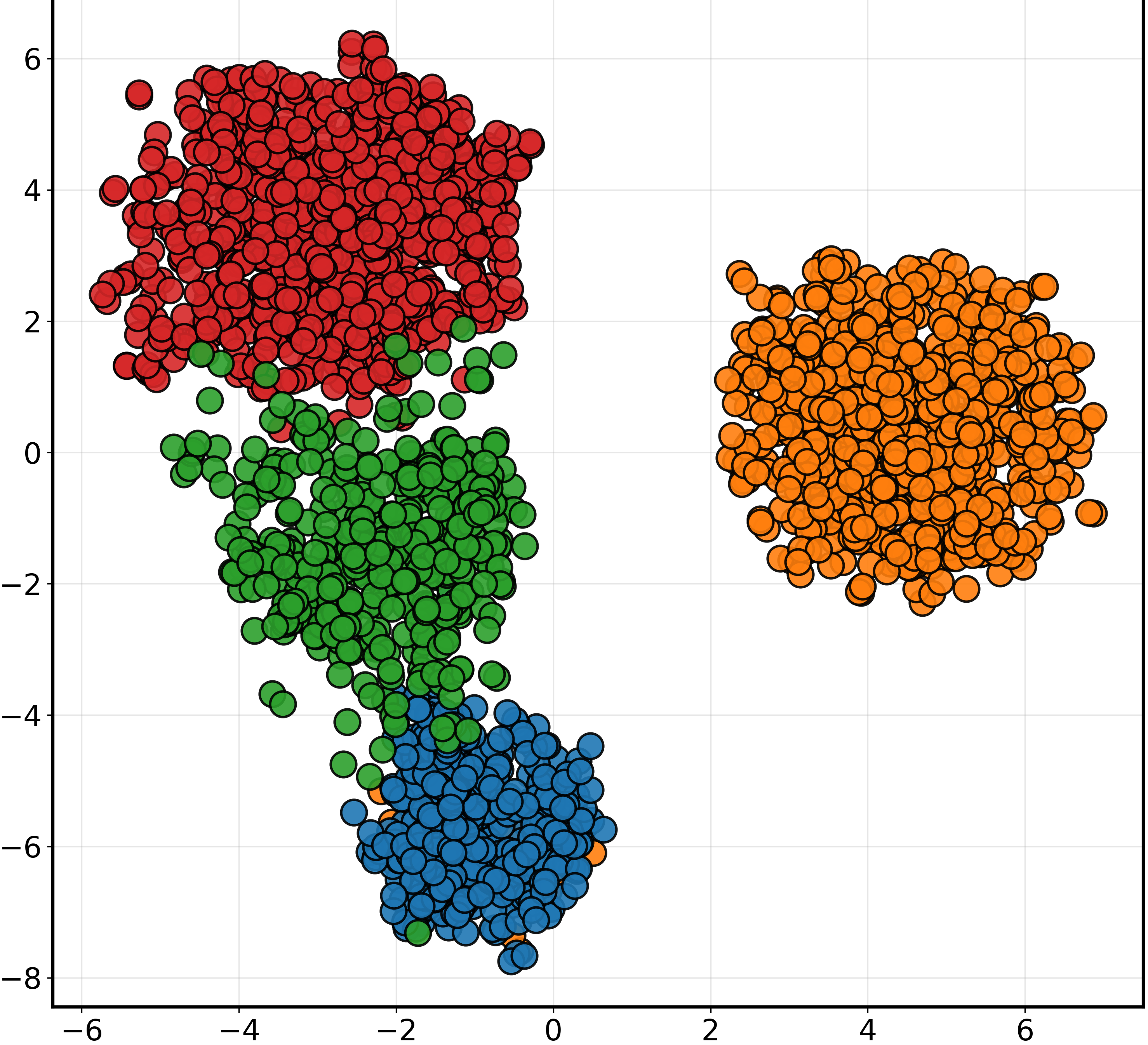}
\label{fig:tsne_no_pcm_ams}
} &
\subfigure[With PCM]{
\includegraphics[width=0.29\linewidth]{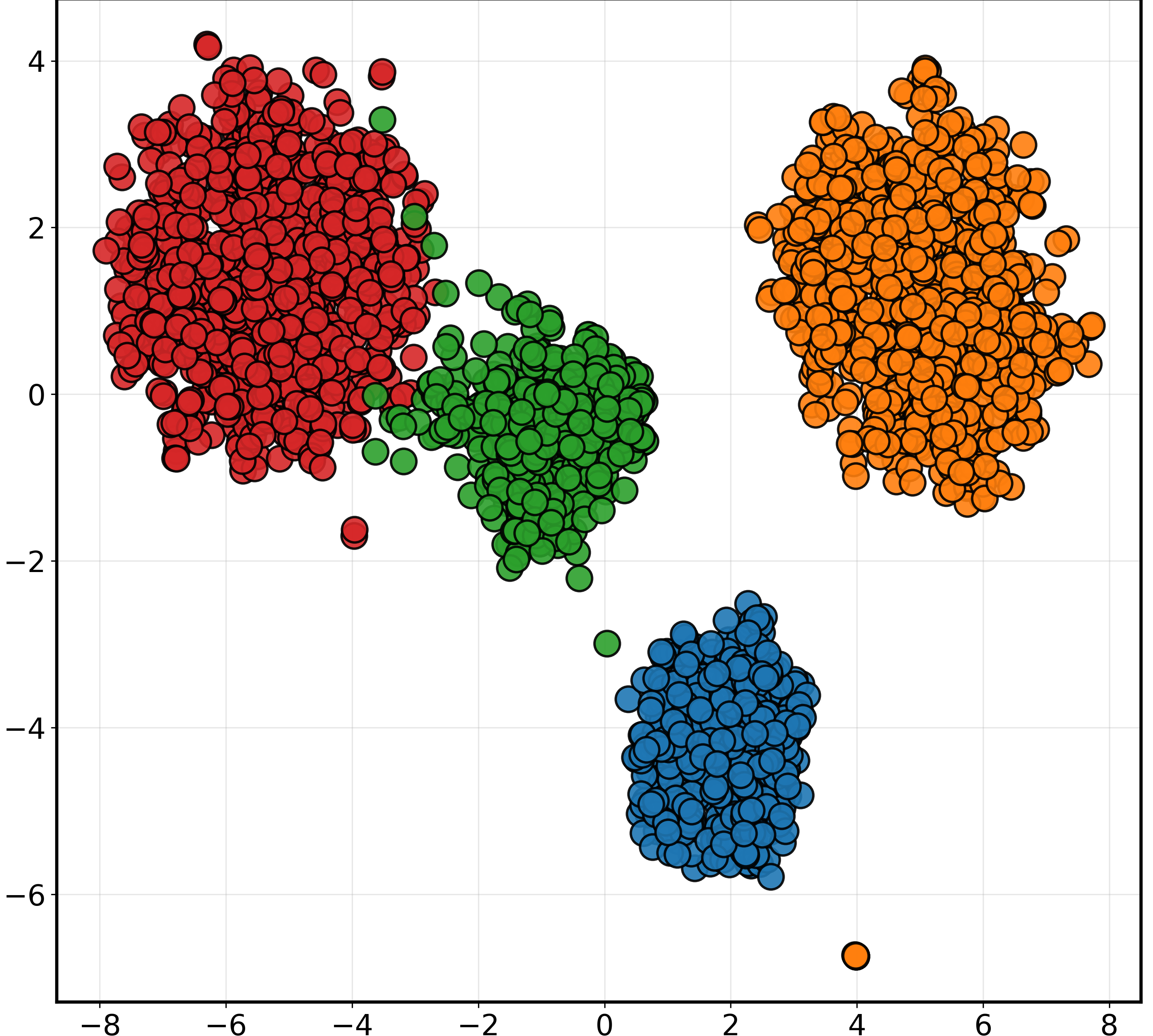}
\label{fig:tsne_with_pcm}
} &
\subfigure[With PCM \& AMS]{
\includegraphics[width=0.29\linewidth]{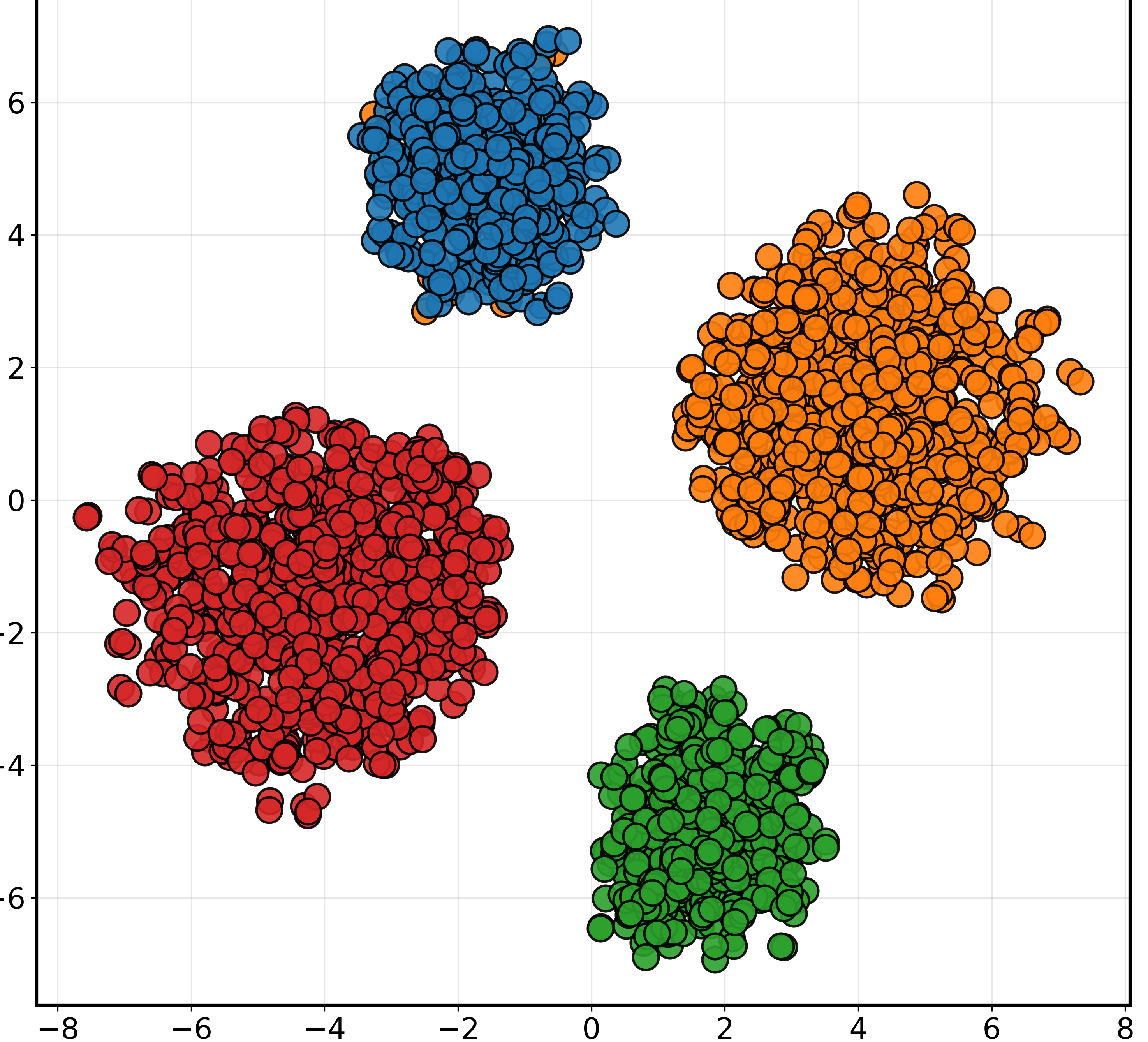}
\label{fig:tsne_with_pcm_ams}
} &
\raisebox{1\height}{\includegraphics[height=0.1\linewidth]{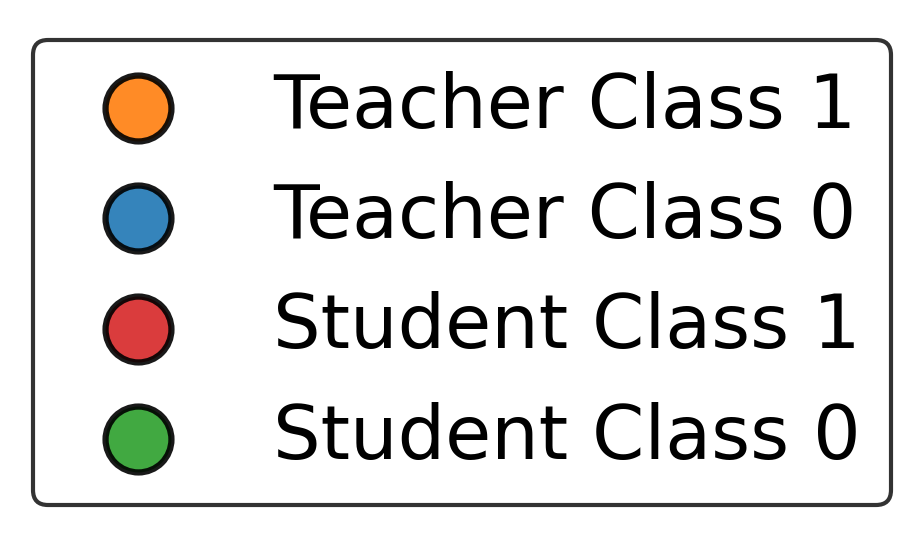}}
\end{tabular}
}
\caption{t-SNE visualization of MRI feature embeddings under different training configurations (50\% missing modality). }
\label{fig:tsne_analysis}
\end{figure}

To evaluate the impact of PGAD's core components on feature discriminability, we visualize the MRI feature embeddings from the student network using t-SNE, under a 50\% missing modality rate. Three configurations are compared: (1) baseline (without PCM or AMS), (2) with PCM only, and (3) the complete PGAD model (with both PCM and AMS), as depicted in Fig.~\ref{fig:tsne_analysis}.

The visualization reveals a clear progression in cluster compactness and class separation. The baseline model (Fig.~\ref{fig:tsne_no_pcm_ams}) exhibits significant class overlap, indicating poor feature discrimination. Incorporating PCM (Fig.~\ref{fig:tsne_with_pcm}) markedly improves class separation, with features forming tighter, more distinct clusters. This demonstrates PCM's effectiveness in aligning unpaired MRI samples with their class prototypes, thereby preserving discriminative information. The complete PGAD model (Fig.~\ref{fig:tsne_with_pcm_ams}) achieves the most optimal clustering, with the highest inter-class separation and intra-class compactness. This visual evidence directly correlates with the quantitative results in Table~\ref{tab:ablation_prototype_source}, where the full model achieves peak performance, confirming that the synergistic integration of PCM and AMS is crucial for learning robust and discriminative features under data incompleteness.

\subsubsection{Biomarker Identification via Grad-CAM}
\begin{figure}[htbp]
    \centering
    \includegraphics[width=0.9\columnwidth]{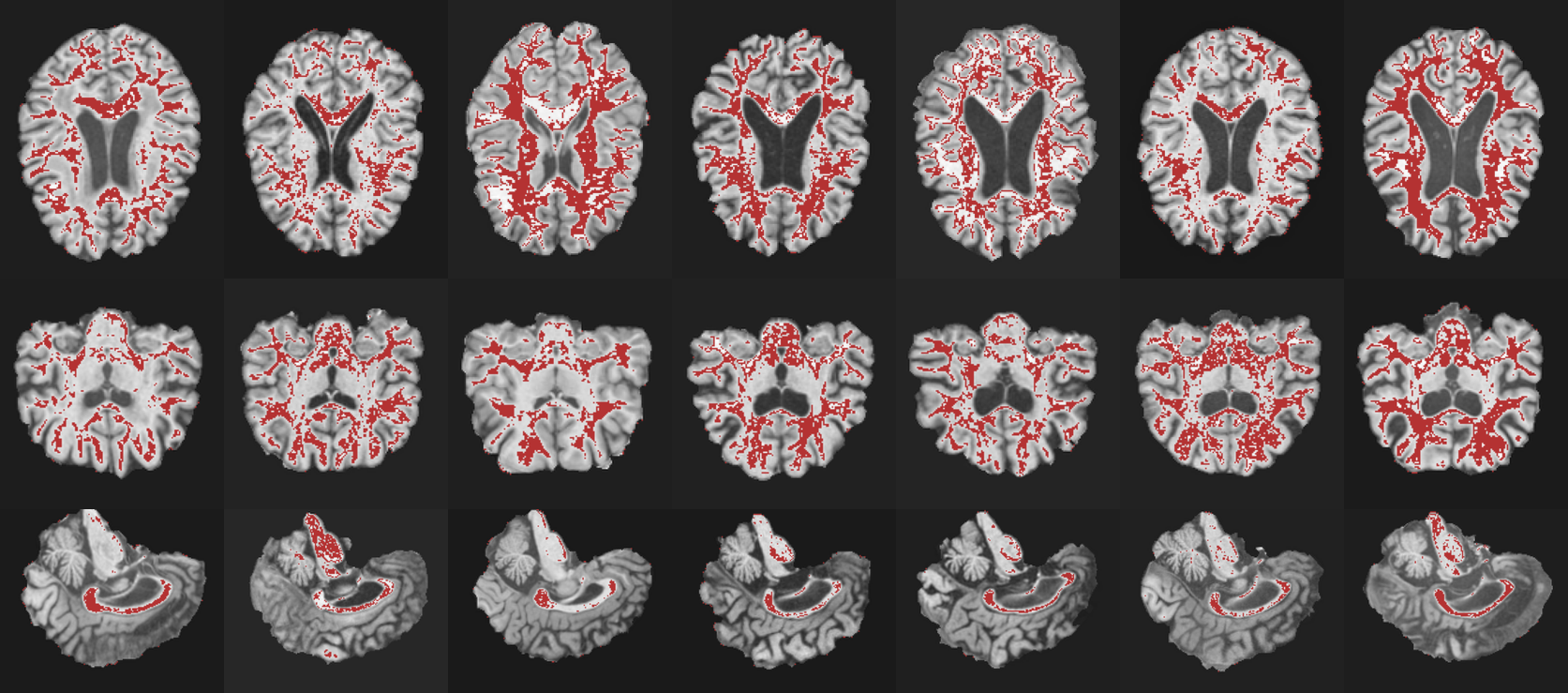}
    \caption{Grad-CAM visualization of the model's attention regions on MRI slices. The red-highlighted areas indicate regions most relevant to AD classification in axial, coronal, and sagittal views.}
    \label{visualization_analysis}
\end{figure}

To assess the clinical relevance of our model, we analyze its attention maps using Grad-CAM on MRI slices, as shown in Fig.~\ref{visualization_analysis}. The results indicate that PGAD consistently focuses on brain regions with established pathological significance in AD.

Specifically, the model highlights areas of the hippocampus, entorhinal cortex, and posterior cingulate cortex, which are known to be among the first affected by AD-related atrophy. It also identifies patterns of cortical thinning in the temporal and parietal lobes, consistent with the disease's progression. These attention patterns exhibit high stability across different subjects and missing modality rates, with a mean Dice coefficient of 0.81 for attention map similarity. This consistency, confirmed by clinical experts, validates that the model learns biologically plausible features rather than spurious correlations. The localization of these known AD biomarkers underscores the model's potential for providing interpretable and trustworthy diagnostic support in clinical practice.

\section{Conclusion}
We introduced the PGAD framework to address the challenges of aligning missing modality features and improving knowledge transfer in Alzheimer’s Disease diagnosis. By integrating Prototype Consistency Matching (PCM) and Adaptive Multi-Modal Sampling (AMS), PGAD effectively leverages incomplete multi-modal data, mitigating feature misalignment and enhancing cross-modal learning. Experimental results demonstrate that PGAD achieves state-of-the-art performance in both AD classification and MCI conversion prediction under various missing rates. Ablation studies confirm the effectiveness of our proposed components. These findings highlight the potential of prototype-guided learning and adaptive sampling for improving multi-modal medical imaging under real-world missing data conditions. Future work will focus on extending the PGAD framework to incorporate three or more modalities and evaluating its performance in more complex, multi-center clinical scenarios.


\end{document}